\documentclass[a4paper]{jpconf}
\usepackage{graphicx}
\begin{document}
\title{VSOP2/\textit{ASTRO-G} Project}

\author{Masato Tsuboi\footnote{Project Scientist, \textit{ASTRO-G} project of ISAS, JAXA}}

\address{Institute of Space and Astronautical Science (ISAS), 
Japan Aerospace Exploration Agency (JAXA), 
3-1-1 Yoshinodai, Sagamihara, Kanagawa, 229-8510
Japan}

\ead{tsuboi@vsop.isas.jaxa.jp}

\begin{abstract}
We introduce a new space VLBI project, the Second VLBI Space Observatory Program (VSOP2),  following the success of the VLBI Space Observatory Program (VSOP1). VSOP2 has 10 times higher angular resolution, up to about 40 micro arcseconds, 10 times higher frequency up to 43 GHz, and 10 times higher sensitivity compared to VSOP1. Then VSOP2 should become a most powerful tool to observe innermost regions of AGN and astronomical masers. \textit{ASTRO-G} is a spacecraft for VSOP2 project constructing in ISAS/JAXA since July 2007. \textit{ASTRO-G} will be launched by JAXA 
H-IIA 
rocket in fiscal year 2012. \textit{ASTRO-G} and ground-based facilities are combined as VSOP2. To achieve the good observation performances, we must realize new technologies. They are large precision antenna, fast-position switching capability, new LNAs, and ultra wide-band down link, etc.. VSOP2 is a huge observation system involving  \textit{ASTRO-G}, ground radio telescopes, tracking stations, and correlators, one institute can not prepare a whole system of VSOP2. Then we must need close international collaboration to get sufficient quality of resultant maps and  to give a sufficient quantity of observation time  for astronomical community.  We  formed a  new international council  to provide guidance on scientific aspects related of  VSOP2, currently called the VSOP2 International Science Council (VISC2).
\end{abstract}

\section{Introduction}
The pioneering history of radio telescopes is in some ways the history of the quest for higher angular resolution.  The angular resolution of  a radio interferometer is given by $\lambda/D$, where $D$ is baseline length or antenna spacing. The upper limit of $D$ was usually much less than 1,000 km for connected interferometers. 
In order to push this limit, VLBI (Very-Long-Baseline Interferometry) was invented. VLBI is made by connecting existing radio telescopes with magnetic tapes instead of cables. Then VLBI observes radio sources with relatively high brightness temperature like AGN and astronomical masers.  Superluminal motion of AGN jet was shown only by very high angular resolution of VLBI. However, ground-based VLBI has an upper limit of angular resolution because $D$ must not be larger than the diameter of the earth, 13,000 km. In order to push this limit, space VLBI was formulated. Space VLBI is an interferometer connecting radio telescopes in orbit and ground-base ones.  Since the baseline length of space VLBI can extend beyond the diameter of the earth, angular resolution of space VLBI is free from this limit.  However, it is very difficult to set a radio telescope in orbit. 

Space VLBI was realized in 1997 as the first space VLBI project, VSOP1 (VLBI Space Observatory Program)[1,2],  after some preceding experiments and vast amounts of effort. The spacecraft for VSOP1 is known as the first dedicated space VLBI satellite, \textit{HALCA}. This was launched by a M-V rocket of Institute of Space and Astronautical Science (ISAS) from Unchinoura Space Center (USC). The angular resolution of VSOP1 is 3 times better than that of ground-based VLBI at 1.6 and 5 GHz (VSOP1 also had 22-GHz band system, which unfortunately suffered a serious loss of sensitivity at the launch). Seven hundred observations of AGN jets were made in the mission lifetime of 7 years. 

Figure 1 shows a schematic display of the comparison between space VLBI and ground-based VLBI.  The Second VLBI Space Observatory Program (VSOP2) is a successor of VSOP1 [3].  \textit{ASTRO-G} is the satellite for VSOP2, which will be launched by ISAS/JAXA in fiscal year 2012. The construction project of \textit{ASTRO-G} has been started since July 2007. This project is a joint-project between ISAS, NAOJ (National Astronomical Observatory, Japan) and the universities. Now, basic design of the satellite and developments of several components for \textit{ASTRO-G} are ongoing.  A review of a basic design of the satellite (PDR: preliminary design review) is scheduled at March of 2009.  After decision of the design (CDR: critical design review) in 2009, we will start the production of the flight model of \textit{ASTRO-G}. 
This satellite will be launched with an H-IIA rocket from Tanegashima Space Center (TNSC) of JAXA. 

VSOP2 has 10 times higher angular resolution, 10 times higher frequency, and 10 times higher sensitivity compared to VSOP1. VSOP2 will be an astronomical tool with unprecedented angular resolution in all wavelength to explore innermost regions of AGN and astronomical masers. The VSOP2 programme foresees a powerful astronomical instrument involving  \textit{ASTRO-G}, ground radio telescopes, tracking stations, and correlators.  For this reason, ISAS alone can not all different aspects of this complex project, which will be made in collaboration with international space and ground-base partners. 

\begin{figure}[h]
\begin{center}
\includegraphics[width=28pc]{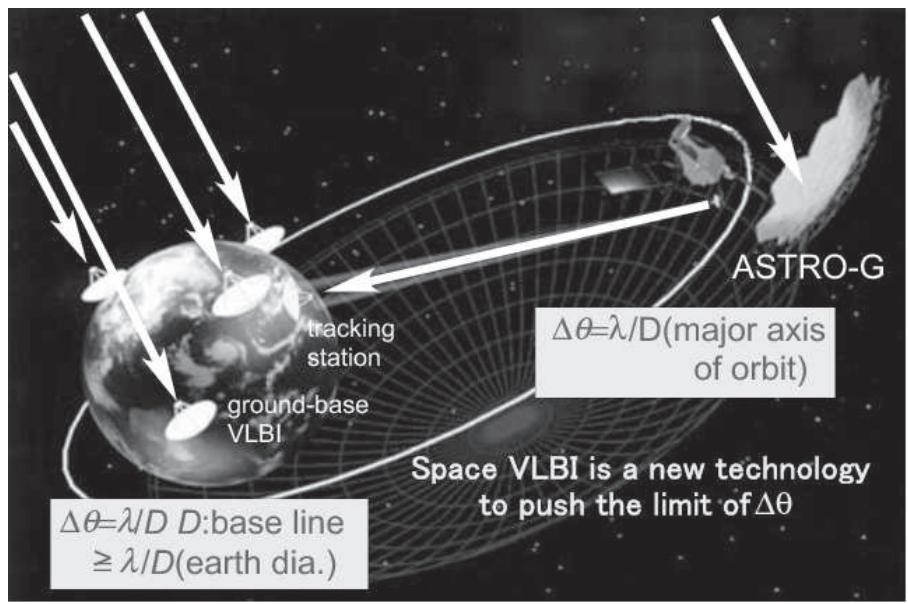}
\hspace{2pc}
\caption{\label{label}Schematic display of the comparison between space VLBI and ground-based VLBI.}
\end{center}
\end{figure}
 
\begin{figure}[h]
\begin{center}
\includegraphics[width=28pc]{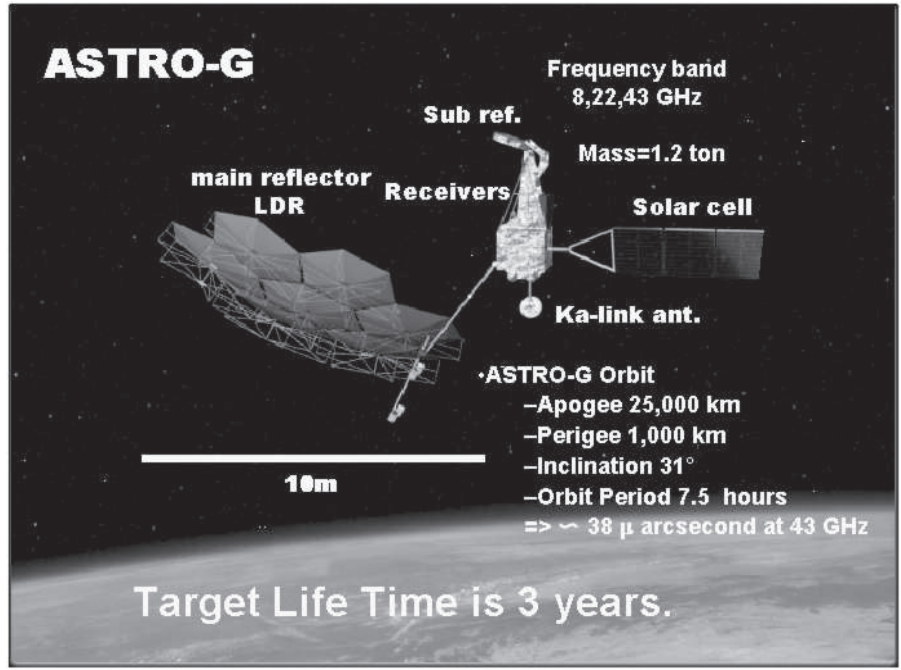}
\hspace{2pc}
\caption{\label{label}Artist's impression of the \textit{ASTRO-G} satellite around perigee.}
\end{center}
\end{figure}

\section{Imaging Capability and Scientific Objectives of VSOP2}
\subsection{Expected Imaging Capability}
Here we describe the expected imaging capability of VSOP2. Figure 2 shows an artist's impression of the \textit{ASTRO-G} satellite around perigee. The total size of the \textit{ASTRO-G} satellite is as large as a tennis coat.
The observing frequency bands of the \textit{ASTRO-G} satellite are 8, 22, and 43 GHz.  There are many large radio telescopes at 8 GHz in the world, being this frequency band used for down-link satellite communication.  The participation of many telescopes in VSOP2 increases its imaging capability. The 22 and 43 GHz bands involve H$_{2}$O ($\nu=22.235$ GHz) and SiO ($\nu=43.122$ and $42.820$  GHz) masers, respectively.  These frequency bands are summarized in Table 1. 

\begin{table}
\centering
\caption{\textit{ASTRO-G} parameters}
\label{} 
\begin{tabular}{lcl}
\hline
    orbit parameters&&\\
\hline
apogee&25,000&km\\
perigee&1,000&km\\
inclination angle & 31$^{\circ}$&\\
orbital period&7.5&hr\\
\hline
frequency band &&\\
\hline
    X &8.0-8.8&GHz\\
    K &20.6-22.6&GHz\\
    Q &41.0-45.0&GHz\\
   
\hline
\end{tabular}
\end{table}

The planned apogee and perigee of the orbit of \textit{ASTRO-G} are 25,000 and 1,000 km, respectively. The orbit provides the maximum baseline length of 35,000 km for VSOP2. The orbital period is about 7.5 hours.  VSOP2 will yield an angular resolution of 38 micro arcseconds at 43 GHz.
 The inclination angle of the orbit is 31$^{\circ}$, which has a influence on the common observation coverage of ground telescopes and the coverage of down link from \textit{ASTRO-G} to data tracking stations. These are important to realize good quality of the resultant maps. The parameters are also summarized in Table 1. 

The expected mission lifetime of \textit{ASTRO-G} satellite is 3 years. This limitation will be caused by the following facts:
\textit{ASTRO-G} has cooled mm-wave receivers by Stirling cycle refrigerator, which will be operated over 3 years. However, the mm-wave receiver has large heat consumption compared to IR detectors. The cooling capacity will decrease to the heat consumption in 3 years.
Furthermore, the radiation condition of \textit{ASTRO-G} is as hard as total dose level of $10^{10}$ rad in 3 years because \textit{ASTRO-G} will pass the radiation belt 3 times a day. This will have an adverse effect on the surface accuracy of the telescope antenna.  The sensitivity of VSOP2  will play an important role in determination of the imaging capability.  In addition, the  imaging capability also depends considerably on the performances of ground radio telescopes. They will be  summarized after the section of the antenna and LNAs.

\subsection{Expected Scientific Objectives}
\subsubsection{Active Galactic Nuclei (AGN)}
The high angular resolution of 38 micro arcsecond at 43 GHz will make possible to study the neighborhood of an AGN black hole in unprecedented quality. Figure 3 shows the comparison among the predicted accretion disk size of the nearest AGN, M\,87, and the synthesized beam sizes of VSOP2, VSOP1, and the VLBA. The synthesized beam sizes of VSOP2 corresponds to $13R_s$, which is smaller than the predicted accretion disk of M\,87. Using VSOP2, we may image the accretion disk of M\,87 if it has a sufficient brightness. 

The resolving an accretion disk disk might solve the following problems about the AGN jet formation.  (i) is the origin of AGN jet in the neighboring region of the black hole or the accretion disk (e.g., [5]).  VSOP2 will make clear from where the jets are ejected.  (ii) is what is the role of magnetic field for acceleration of AGN jet (e.g., [6]). VSOP2 can observe polarization at multi-frequencies. Then VSOP2 will make clear the magnetic field structure in the base of AGN compensating Faraday effect (cf. [7]). (iii) is acceleration efficiencies of the jet. VSOP2 will make clear the energy distribution of the relativistic electrons in the base of AGN from multi-frequency observations.

Blazars are also key sources for exploring the nature and physics of AGN jets. VSOP1 could image blazars at 1-pc scale. VSOP1 had provided a great progress in VLBI astronomy of blazars (e.g.,  [8]). On the other hand, X-ray observations estimate from the timescales of X-ray flares the size of the emission region to be about 0.1pc or less. The high-resolution observations of VSOP2 will resolve the gap between these values. Neighborhood blazars, for example, Mrk421 and Mrk501 are nice candidates for this study. 

\begin{figure}[h]
\includegraphics[width=20pc]{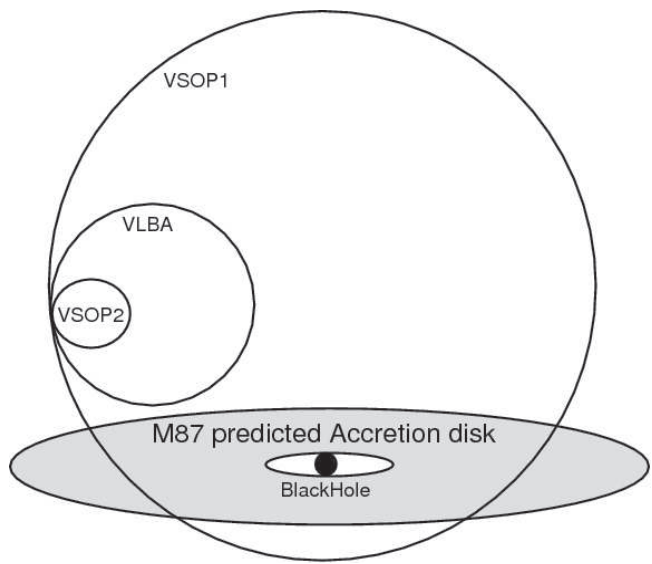}
\hfill
\begin{minipage}[b]{14pc}
\caption{\label{label}Comparison among the predicted accretion disk size of the nearest AGN, M\,87, and the synthesized beam sizes of VSOP2 (38$\mu as$), VSOP1 (400$\mu as$), and VLBA (100$\mu as$) at selected frequencies.}
\end{minipage}
\end{figure}

\subsubsection{Young Stellar Objects}
Young stellar objects  (YSO) are as well promising targets of VSOP2. The angular resolution of VSOP-2 is comparable to the radius of a YSO. Unfortunately, VSOP2 cannot observe thermal emission from gas in/around YSO. However, motions of H$_2$O and/or SiO maser spots can trace structure and kinematics of such gas, or position and transversal velocity, and  acceleration of such gas (e.g., [9]).  Then, VSOP2 will make clear how  such gas accretes into the YSO and  how the angular momentum is exchanged.  

And we already knew that large flares occur around the surface of YSO (see Figure 4).  Direct imaging of these flares may be a fantastic target of VSOP2. If imaging is possible, we can prove a flare scaling law, which is expected widely from solar micro-flares to YSOs (e.g., [10]).  

\subsubsection{Cosmology}
The H$_2$O masers of NGC4258 is a famous example of disk masers resolved by VLBI. Cosmology using disk masers of galaxies may be another scientific target of VSOP2. Acceleration of central maser component, $a$, is given by $a=v^2/r$, where $v$ is rotation velocity of maser component and $r$ is the radius of the disk.  On the other hand, $r$ is also given by $r=d\theta$, where $d$ is the distance of disk maser and $\theta$ is the angular radius of disk maser. Then $d$ is given by a simple formula, $d=v^2/a\theta$. VSOP2 observations will provide the values of $a$, $v$, and $\theta$. We can determine $d$ based on VSOP2 only. Figure 5 shows a schematic display of the measurement of the distance to maser disk of AGN using VSOP2. VSOP2 can observe several galaxies hosting strong H$_2$O masers. 
VSOP2 can measure the Hubble constant with high precision, because the systemic velocities of galaxies are precisely measured with optical observations.  Precise values of Hubble constant and its change for redshift will provide additional information about the dark energy content of the universe. 

\begin{figure}[h]
\includegraphics[width=22pc]{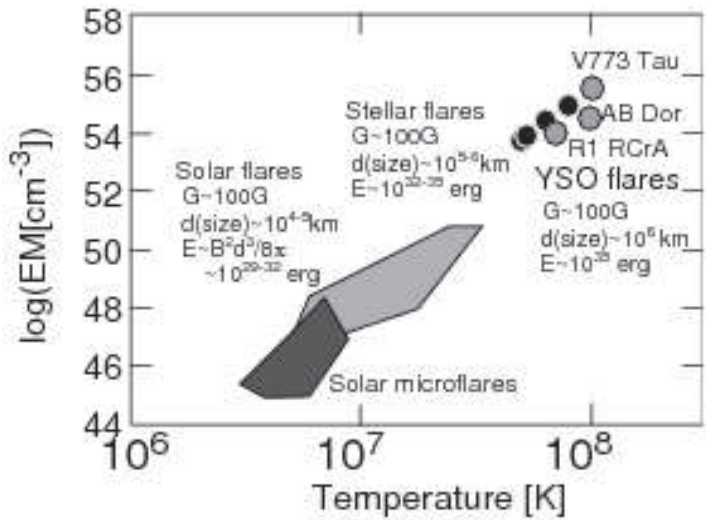}
\hfill
\begin{minipage}[b]{14pc}
\caption{\label{label}Comparison among the size, energy, magnetic field strength of flares occured at Sun, stars, and YSO, overlaid on the plane of the emission measure and temperature of the flares (Shibata and Yokoyama 1999).}
\end{minipage}
\end{figure}

\begin{figure}[h]
\begin{center}
\includegraphics[width=28pc]{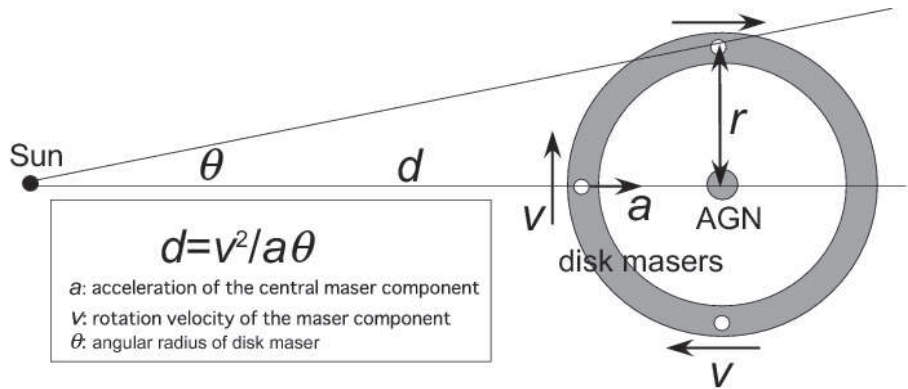}
\hspace{2pc}
\caption{\label{label} Schematic display of the measurement of the distance to a maser disk of AGN using VSOP2.}
\end{center}
\end{figure}

\begin{figure}[h]
\begin{center}
\includegraphics[width=28pc]{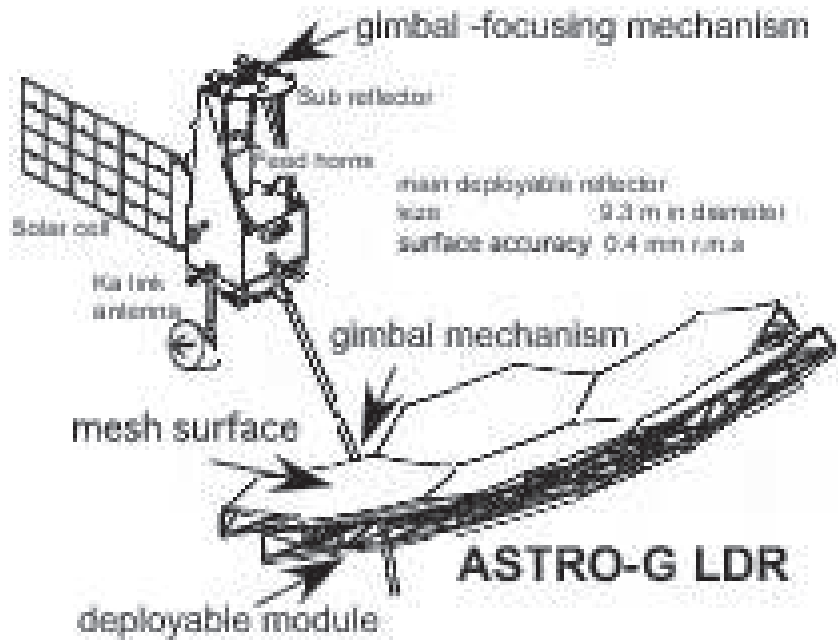}
\hspace{2pc}
\caption{\label{label} Schematic display of the space radio telescope of \textit{ASTRO-G} satellite.}
\end{center}
\end{figure}

\section{The Space Radio Telescope of \textit{ASTRO-G} }
The space radio telescope of \textit{ASTRO-G} satellite is an offset Cassegrain-type telescope which has a large precision antenna, a control momentum gyro for fast-position switching capability, LNA receivers, and  ultra wide-band down link system [11].

\subsection{Large Deployable Reflector}
Figure 6 shows schematically the space radio telescope of \textit{ASTRO-G} satellite. The antenna consists of a 9.3-m paraboloid main reflector (Large Deployable Reflector, hereafter LDR), a hyperboloid sub-reflector, and three feed horns at 8, 22, 43 GHz. 

LDR has 7 hexagonal metal mesh-surface modules, which are deployed on the orbit [12]. Radial rib-hoop cable structure is newly adopted for the modules to form a surface with accuracy of 0.4 mm-rms.  The surface accuracy of \textit{ASTRO-G} LDR is 10 times higher than that of the large deployable main reflector of \textit{ETS VIII}, which is an engineering satellite of JAXA launched in 2006. The total weight of LDR is about 200 kg. We will make one-module as an engineering model (EM) up to spring 2009 to evaluate the deployable mechanism and the accuracy.  We can focus LDR on the orbit using gimbal -focusing mechanism to compensate the large-scale deformation of LDR. 

The feed horns at 8, 22, 43 GHz are multi-mode horns for small sizing on the focal plane. We are now designing the antenna optics with simulation based on physical optics method. Ideal antenna efficiencies, depending only on the optics of the LDR, are estimated to be $\eta_0=64-68 \%$ for a $-10$ dB edge level of the main reflector. The antenna optical performances at the present are  listed in Table 2. The beam patterns have good axial symmetry and good cross-polarization level as low as $< -25$ dB. 

The metal mesh surface is a key component for the light weight of the LDR. This is woven mesh of gold plated molybdenum fiber. The real antenna RF performances of the LDR suffer from degradations in the surface accuracy and the reflection efficiency, $\eta_S$, and the instrumental polarization of the metal mesh surface.  Because it is difficult to calculate the RF performances of the metal mesh surface itself, we are measuring the reflection efficiency and the instrumental polarization as functions of the tension of the metal mesh surface with a test apparatus. These measured values are also need for optimum design of LDR back structures.

Then, the overall expected antenna efficiency is given by $$\eta_A=\eta_0\eta_S\exp{\{-(4\pi\epsilon/\lambda)^2\}},$$ where $\eta_0$ is the ideal antenna efficiencies depending only on the illumination pattern and the spillover of the LDR and $\eta_S$ is the reflection efficiency of the metal mesh surface.  

The values of $\eta_A$ are also listed in Table 2. The last column shows best estimates of the aperture efficiency at present. We hope dramatic improvement even on the orbit compared to the \textit{HALCA} antenna. We will be able to achieve the performance at least at the begin of the life time. However, the radiation condition of \textit{ASTRO-G} is very hard because it will pass radiation belt over 3000 times during the mission life time. The performances at the end of the life time are now discussing.

\begin{table}
\centering
\caption{Expected antenna performances of LDR}
\label{LDR1} 
\begin{tabular}{ccccccc}
\hline           
Band&Gain&$\eta_0$&x-pol.&$\eta_S$&$\exp{\{-(4\pi\epsilon/\lambda)^2\}}$&$\eta_A$\\

[GHz]&[dB]&\ &[dB]&\ &\ &\ \\
\hline
    $8.0-8.8$&$56.3$&$0.64$&$-27.8$&$1.00$&$0.98$&$0.60$\\
$20.6-22.6$&$64.9$&$0.67$&$-32.9$&$1.00$&$0.88$&$0.58$\\
$41.0-45.0$&$70.8$&$0.68$&$-35.6$&$0.87$&$0.61$&$0.35$\\
\hline
\end{tabular}
\end{table}

\subsection{Fast Position-switching Capability}
The \textit{ASTRO-G} satellite has fast position-switching capability for phase referencing observations.  Although a space telescope is free from atmospheric effects, VSOP2 is not completely so because VSOP2 is an interferometer between \textit{ASTRO-G} and ground based telescopes. We need to slew \textit{ASTRO-G} from a target to a calibrator within 3 degrees in 15 seconds to compensate for the atmospheric effects.  Then, we would use Control Momentum Gyro (CMG) technology [13], which has been used in the \textit{Space Shuttle}. 

\begin{figure}[h]
\begin{center}
\includegraphics[width=20pc]{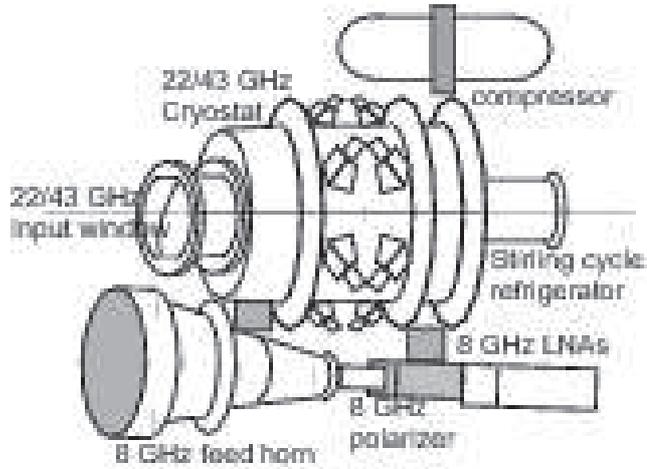}
\hspace{2pc}
\caption{\label{label} Schematic display of the 8, 22, and 43 GHz band receivers of the \textit{ASTRO-G} satellite.}
\end{center}
\end{figure}

\begin{figure}[h]
\includegraphics[width=22pc]{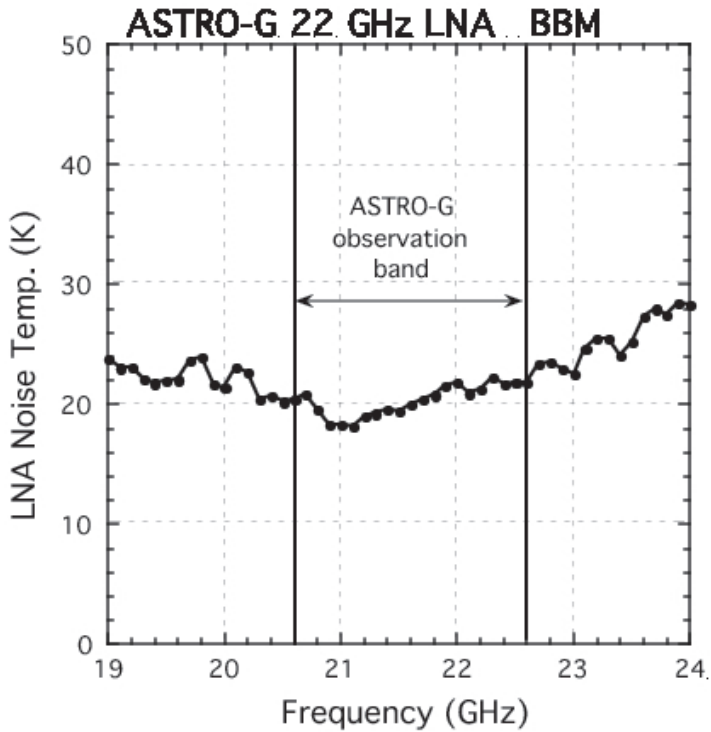}
\hfill
\begin{minipage}[b]{14pc}
\caption{\label{label}Present status of the 22 GHz LNA of \textit{ASTRO-G}. This LNA is at the BBM phase. The LNA noise temperature is as low as the technical goal demonstrated in the project book of \textit{ASTRO-G}.}
\end{minipage}
\end{figure}

\subsection{Receiver System}
Figure 7 shows the 8, 22, 43 GHz band receivers of \textit{ASTRO-G}. The low noise amplifier (LNA) is the most important component of the receiver. LNAs used in space must be ultra-low noise and un-conditionally stable.  The LNAs at 22 and 43 GHz are cooled to 30 K and the dual circular polarization feed horns are cooled to 100 K by a Stirling-cycle refrigerator. We use GaAs monolithic microwave integrated circuits (GaAs MMIC) technology for these LNAs. On the other hand, the LNA at 8 GHz is located at the radiator panel of the receiver system. This is cooled passively. The feed horns are dual circular polarization feeds which are  juxtaposed on the Cassegrain focal plane of LDR. 

The noise temperatures of LNAs at the bread board model  (BBM)  phase are 60 K at 8 GHz, 20 K at 22 GHz, and 35 K at 43 GHz. Figure 8 shows the present status of the noise temperatures of the 22 GHz LNA at the BBM phase. The LNA at 43 GHz does not yet achieve the target values expected in the \textit{ASTRO-G} project book. We measured (estimated for some components) loss and noise of the receiver components and  estimate system noise temperatures at the present.  We assumed that the system equivalent flux density ($SEFD$) is given by $$SEFD=2kT_{sys}/A_e.$$ Table 3 shows the SEFDs of VSOP2 expected for three receivers.  Numbers in parentheses are target values at the project book. The present estimated figures do not achieve the target values. Because there are some possibility of overestimation for some parts, they may be the upper limit for SEFDs. We will make receivers at the engineering model (EM) phase in autumn 2008; the performances of the EM receivers will be reported in the beginning of 2009. 

\begin{table}
\centering
\caption{Expected Performances of VSOP2 (Target Values in Brackets)}
\hspace{2pc}
\label{ASTRO-G1} 
\begin{tabular}{cccc}
\hline
 Band&resolution&SEFD&7-$\sigma$detection with VLBA\\
 
  [GHz]&[$\mu$as]&[mJy]&[mJy]\\
\hline
    $8$&$205$&$6100 (4080)$&$32(23)$\\
$22$&$75$&$3600 (2200)$&$72(50)$\\
$43$&$38$&$7550 (3170)$&$188(107)$\\
\hline
\end{tabular}
\end{table}

\subsection{IF and  Link System}
We can observe simultaneously dual polarizations at each frequency. However, we can not observe two frequencies at once because different frequency bands use the different feed horns. We can switch two frequencies at least within one minute. The bandwidth and the bit levels of the analog-to-digital converter (ADC) of \textit{ASTRO-G} are 128 MHz and 2 bit, or 256 MHz and 1 bit.
Ground-based VLBI radio telescopes record the signal to recording media, tape, disk, etc. In contrast, \textit{ASTRO-G} satellite has no recording media.  It sends directly the VLBI-formatted data to the tracking stations. The data are sent through a broadband down link of $ band width\sim900$ MHz at 37.5 GHz with a bit rate of 1 Gbps quadrature phase shift keying (QPSK). Simultaneously, the phase transfer carrier signal to the local oscillator system of the \textit{ASTRO-G} is sent through the up-link at 40 GHz. The ultra-wide band down-link may be also a technical challenge. Figure 9 shows a block diagram of the observation and data link systems of the \textit{ASTRO-G} satellite.
We plan at least 3 tracking stations in the world for obtaining sufficient observation time. However, ISAS has only one dedicated VLBI tracking station at Usuda, Japan. International collaboration is essential for the success of VSOP2. Additionally, a tracking station in southern hemisphere is important for keeping a high observation efficiency of VSOP2.

\begin{figure}[h]
\begin{center}
\includegraphics[width=28pc]{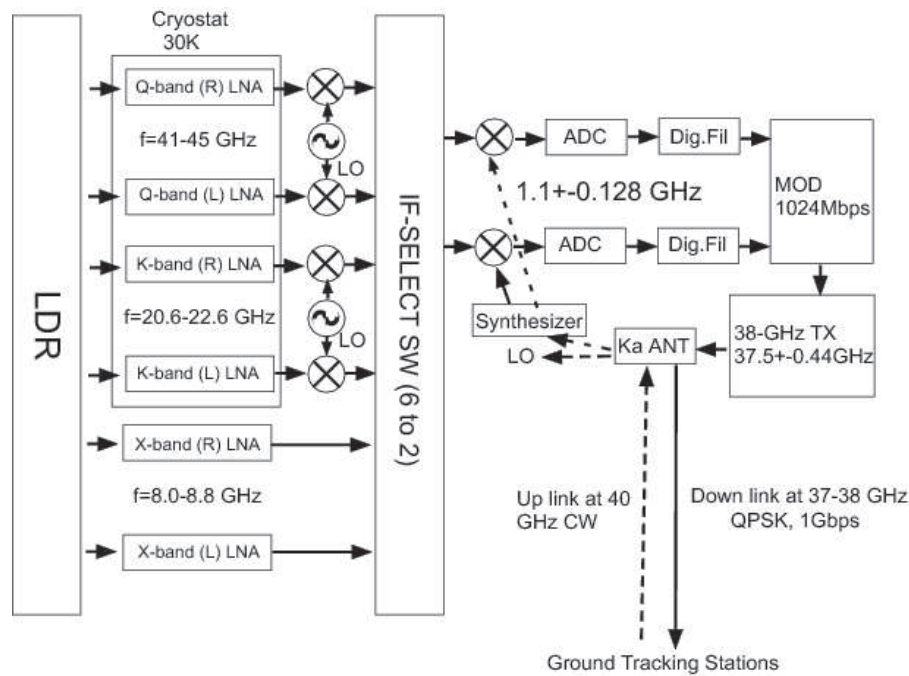}
\hspace{2pc}
\caption{\label{label} Block diagram of observation and data link systems of \textit{ASTRO-G} satellite}
\end{center}
\end{figure}

\subsection{High Accuracy Orbit Determination}
\textit{ASTRO-G} needs a very high accuracy orbit determination for astrometry.
We use both satellite laser ranging (SLR) and GPS systems for a high accuracy orbit determination.  The accuracy of the orbit determination is required to be at least 10 cm to get any advantage against ground-base VLBI.  The very high accuracy determination using GPS may be also technical challenge.  The altitude of \textit{ASTRO-G} around the apogee is much higher than that of a GPS constellation. Since the antenna of GPS satellites is usually pointed to earth, \textit{ASTRO-G} around the apogee only receives the signal from the GPS satellites located on the far side of earth. It is difficult to receive the signals from over four satellites, which are required to determine the orbit of \textit{ASTRO-G} directly. \textit{ASTRO-G} also has a corner cube reflector for satellite laser ranging at the earth-pointing system of Ka-band link antenna.  

\section{VSOP2 International Science Council (VISC2) Formation}
As mentioned previously, we need a world-wide collaboration for VSOP2 both in space and ground telescopes to make resultant maps with sufficient quality and  to give a sufficient quantity of observation time  for  international astronomical community.  However, any international collaboration is a complicated issue.  VSOP1 formed the VSOP International Science Council (VISC1) as an international body to provide guidance on scientific aspects related of the mission.  We are acknowledging that VISC1 functioned successfully in maximizing scientific result with VSOP1.  In a similar fashion for VSOP2, we formed VISC2.  The primary function of VISC2 will be to form an international consensus about issues relevant to science operation of the VSOP2 mission. VISC2 expects to have face-to-face meeting approximately twice per year and more frequently teleconferences. 

\ack
This work is based on cooperative activity of the \textit{ASTRO-G} project team.

\section*{References}

\end{document}